\documentclass[a4paper,11pt]{article}
\usepackage{graphicx}
\usepackage{dcolumn}
\usepackage{bm}
\usepackage{graphicx,amssymb,amsmath,amsfonts,epsfig}
\usepackage[usenames,dvipsnames]{color}

\title{CFT and Logarithmic Corrections to the Black Hole Entropy Product Formula}
\author{Parthapratim Pradhan\footnote{pppradhan77@gmail.com}  \\ 
\textit{ Department of Physics} \\
\textit{Hiralal Mazumdar Memorial College For Women}\\
\textit{Dakshineswar, Kolkata-700035,  India.}}
\date{}

\begin{document}

\maketitle
\begin{abstract}
We examine the logarithmic corrections to the black hole (BH) entropy product formula of outer horizon and 
inner horizon by taking into account the \emph{effects of statistical quantum fluctuations around the thermal equilibirium} 
and \emph{via conformal field theory} (CFT). We  argue that logarithmic corrections to the BH entropy product formula when 
calculated using CFT and taking into the effects of quantum fluctuations around the thermal equilibirium, the formula  
should \emph{not be universal} and it should also \emph{not be quantized}. These results have been explicitly checked by 
giving several examples. 
\end{abstract}


\maketitle


\section{Introduction}
There has been considerable ongoing excitement in physics of BH thermodynamic product formula 
[particularly area ( or entropy) product formula ] of inner horizon (${\cal H}^{-}$) and outer horizon (${\cal H}^{+}$) 
for a wide variety of BHs~\cite{ah09,cgp11,castro12,mv13,val13,pp14,jh,horava15,grg16} which 
have been examined so far \emph{with out considering any logarithmic correction}. For several cases, the product is 
mass-independent (universal) and in some specific cases the product is not mass-independent. This investigation has been more 
sparked by Cveti\v{c} et al.~\cite{cgp11} for supersymmetric BPS (Bogomol'ni-Prasad-Sommerfield) class of BHs which have  
inner and outer BH entropy of the form ${\cal S}_{\pm} =2\pi \left(\sqrt{N_{L}}\pm\sqrt{N_{R}} \right)$,
where $N_{L}$ and $N_{R}$ are excitation numbers of the left and right moving sectors of a weakly coupled two dimensional (2D) 
CFT. Therefore their product ${\cal S}_{+}{\cal S}_{-}$ \emph{ is quantized in nature by its own right}.

It has been suggested by Larsen~\cite{finn} that BH event horizon is quantized in 
Planck units so it is natural to be valid for Cauchy horizon also. This also indicates the product of area (or entropy ) is 
quantized in terms of quantized charges and quantized angular momentum etc. But this has been discussed without considering any 
logarithmic correction to the BH entropy. Now if we take into account the logarithmic correction to the BH entropy what happens 
in case of logarithmic correction to the BH entropy product formula? What is the quantization rules for logarithmic corrected 
BH entropy product formula? These are the main issues that we will be discussed in this work.

It should be noted that this is the continuation of our earlier investigation~\cite{grg1}. In the previous work, we derived 
the general logarithmic correction to the entropy product formula of  event horizon  and Cauchy horizon  
for various spherically symmetric and axisymmetric BHs by taking into account of the effects of quantum  fluctuations 
around the thermal equilibirium. These corrections are evaluated in terms of some BH thermodynamic parameters namely the 
\emph{specific heat $C_{\pm}$} of ${\cal H}^{\pm}$  and \emph{BH temperature $T_{\pm}$} of ${\cal H}^{\pm}$ respectively.

The logarithmic correction of BH entropy of ${\cal H}^{\pm}$ is described by the key formula  
\begin{eqnarray}
{\cal S}_{\pm}  &=&  \ln \rho_{\pm} ={\cal S}_{0, \pm}-\frac{1}{2} \ln \left|C_{\pm} T_{\pm}^2\right|+...  ~ \label{eq1}
\end{eqnarray}
where $\rho_{\pm}$ is density of states of ${\cal H}^{\pm}$ and ${\cal S}_{\pm}$ is entropy of ${\cal H}^{\pm}$
and their product is derived to be
$$
{\cal S}_{+} {\cal S}_{-}  =
{\cal S}_{0, +}{\cal S}_{0, -}-\frac{1}{2}\left[{\cal S}_{0, +} \ln \left|C_{-} T_{-}^2\right|+{\cal S}_{0,-}
\ln \left|C_{+} T_{+}^2\right|\right]
$$
\begin{eqnarray}
+\frac{1}{4} \ln \left|C_{+} T_{+}^2\right| \ln \left|C_{-} T_{-}^2\right| +...   \label{eq2}
\end{eqnarray}
where ${\cal S}_{0, \pm}$ is the entropy of ${\cal H}^{\pm}$ without logarithmic correction.

In the present work, we shall  compute the general logarithmic correction to BH entropy of ${\cal H}^{\pm}$ whenever we 
have taken \emph{the effects of statistical quantum  fluctuations around the thermal equilibirium}  and by  using exact entropy 
function ${\cal S}_{\pm}(\beta_{\pm})$ according to the formalism borrowed from the quantum theory of gravity 
~\cite{bk72,bk73,bcw73,davies,sommer,vafa,ajak,carlip,carlip1,km,km1,skm,psm,guica}. Whenever, we incorporated the effects of 
quantum  fluctuations around the thermal equilibirium  the Bekenstein-Hawking entropy formula must be corrected and the entropy 
product formula of ${\cal H}^{\pm}$ must also  be corrected. This is the main motivation behind this work. 

It has been known that BHs in Einstein's gravity as well as  other theories of gravity have much larger than the 
Planck scale length where the Bekenstein-Hawking entropy is precisely proportional to the horizon area 
~\cite{bk72,bk73,ajak,carlip,carlip1}. Thus it is quite natural to investigate what the \emph{leading order corrections in  
Bekenstein-Hawking entropy} as well as in \emph{Bekenstein-Hawking entropy product formula} of ${\cal H}^{\pm}$ when one 
can reduce the size of BH. For large BHs, it has been proved that the logarithm of the density of states is exactly the 
Bekenstein-Hawking entropy plus the corrections term $-\chi \ln A$, where $\chi=\frac{3}{2}$ and $A$ is the area of the 
event horizon~\cite{km1,skm}. Thus, it appears that the logarithmic corrections to the Bekenstein-Hawking entropy 
as well as \emph{Bekenstein-Hawking entropy product formula} of ${\cal H}^{\pm}$ is a generic feature of BHs. It has 
been verified earlier by Das et al.~\cite{psm} for ${\cal H}^{+}$ and here we have tried to examine for ${\cal H}^{-}$ 
followed by our earlier investigation~\cite{grg1}. 

It should be emphasized that logarithmic corrections to the Bekenstein-Hawking formula are very interesting and a great 
deal about such corrections is known in string theory and beyond. Logarithmic corrections arise from various sources the 
simplest of which are the statistical fluctuations around thermal equilibirium. These are always present because they arise 
from saddle point Gaussian corrections to the integral that computes the density of states from the partition function. 
In some cases, such as the BTZ BH in pure 3D gravity, these are the only logarithmic corrections to the Bekenstein-Hawking 
entropy. However, more generally the logarithm of the partition function, $ln Z$, itself receives corrections from the 
massless spectrum of particles in the theory whose solution contains the BH. These corrections therefore cannot be determined 
from the BH solution only. They are universal only in the sense that they are independent from the UV completion of the 
theory (See review article~\cite{as14}).

Moreover, it must be noted that a given theory of quantum gravity which will assign a Hilbert space to the ${\cal H}^{+}$ 
such that counting the number of microstates of the Hilbert space which gives us the entropy of ${\cal H}^{+}$ of the 
BH by the Boltzmann's entropy formula 
\begin{eqnarray}
{\cal S}_{+} &=& k_{B} \, \ln \Omega_{+}  \label{kb}
\end{eqnarray}
where $K_{B}$ is the Boltzmann constant and $\Omega_{+}$ is the microstates of ${\cal H}^{+}$ only. Analogously, 
there must exists \emph{inner Hilbert space} for ${\cal H}^{-}$. Therefore the Boltzmann's entropy formula for  
${\cal H}^{-}$ becomes
\begin{eqnarray}
{\cal S}_{-} &=& k_{B}\, \ln \Omega_{-}  \label{kb1}
\end{eqnarray}
where $\Omega_{-}$ is the microstates of ${\cal H}^{-}$. Finally, their product should be 
\begin{eqnarray}
{\cal S}_{+} {\cal S}_{-}  &=& k_{B}^2 \, \ln \Omega_{+} \ln \Omega_{-} ~. \label{kb2}
\end{eqnarray}
Now we turn to  compute the logarithmic corrections to BH entropy product formula by using the
CFT formalism.

\section{Logarithmic Corrections to the BH Entropy Product Formula of  ${\cal H}^{\pm}$ via CFT }
We have started with the partition function~\cite{psm,grg1} of any thermodynamic system consisting ${\cal H}^{\pm}$ should 
read off
\begin{eqnarray}
{\cal Z}_{\pm}(\beta_{\pm}) &=& \int_{0}^{\infty} \rho_{\pm}(E) e^{-\beta_{\pm} E} dE  
~ \label{pf}
\end{eqnarray}
where $T_{\pm}=\frac{1}{\beta_{\pm}}$ can be defined as the temperature of ${\cal H}^{\pm}$. 
We have to set Boltzmann constant $k_{B}=1$.

The density of states of the said thermodynamic system may be expressed by taking an inverse Laplace transformation 
(keeping $E$ fixed) of the partition function defined at ${\cal H}^{\pm}$~\cite{bohr,bhadu,psm}
\begin{eqnarray}
\rho_{\pm}(E) &=& \frac{1}{2 \pi i}\int_{c-i\infty}^{c+i\infty} {\cal Z}_{\pm}(\beta_{\pm}) e^{\beta_{\pm} E} d\beta_{\pm} 
\nonumber\\ 
&=& \frac{1}{2 \pi i}\int_{c-i\infty}^{c+i\infty} e^{S_{\pm}(\beta_{\pm})} d{\beta_{\pm}}~ \label{ipf}
\end{eqnarray}
where $c$ is a real constant and defining
\begin{eqnarray}
S_{\pm} &=&  \ln {\cal Z}_{\pm} +\beta_{\pm} E  ~ \label{ents}
\end{eqnarray}
is the exact entropy of ${\cal H}^{\pm}$ as a function of temperature.

Now if we consider the system to be in equilibirium then the inverse temperature is defined to be 
$\beta_{\pm}=\beta_{0, \pm}$ therefore we can expand the entropy function of ${\cal H}^{\pm}$ as
\begin{eqnarray}
S_{\pm}(\beta_{\pm}) &=&  S_{0, \pm}+\frac{1}{2} (\beta_{\pm}-\beta_{0, \pm})^2 S_{0, \pm}'' + ...  ~\label{sbs0}
\end{eqnarray}
where $S_{0, \pm}: =S_{\pm}(\beta_{0, \pm})$ and $S_{0, \pm}''=\frac{\partial^2 S_{\pm}}{\partial \beta_{\pm}^2}$
at $\beta_{\pm}=\beta_{0, \pm}$.

Revert back Eq. (\ref{sbs0}) in Eq. (\ref{ipf}), we find
\begin{eqnarray}
\rho_{\pm}(E) &=& \frac{e^{S_{0, \pm}}}{2 \pi i}\int_{c-i\infty}^{c+i\infty}
e^\frac{\left(\beta_{\pm}-\beta_{0, \pm}\right)^2 S_{0, \pm}''}{2} d{\beta_{\pm}}~.\label{rhoe}
\end{eqnarray}

Let us put $\beta_{\pm}-\beta_{0, \pm} =i x_{\pm}$ and choosing $c=\beta_{0, \pm}$, $x_{\pm}$ is a real variable 
and evaluating a contour integration one obtains
\begin{eqnarray}
 \rho_{\pm}(E) &=& \frac{e^{S_{0, \pm}}}{\sqrt{2 \pi S_{0, \pm}''}}~.\label{ps0}
\end{eqnarray}

Thus the logarithm of the density of states  gives the corrected entropy of ${\cal H}^{\pm}$:
\begin{eqnarray}
{\cal S}_{\pm}:  &=&  \ln \rho_{\pm} ={\cal S}_{0, \pm}-\frac{1}{2} \ln S_{0, \pm}''+ ... ~. \label{eq3}
\end{eqnarray}
The main aim of this work is to  compute the term $S_{0, \pm}''$ by using the exact entropy function 
${\cal S}_{\pm}(\beta_{\pm})$, evaluated at the eqilibrium temperature $\beta_{\pm}=\beta_{0, \pm}$. 

Assume the exact entropy function ${\cal S}_{\pm}(\beta_{\pm})$ of ${\cal H}^{\pm}$ \footnote{The entropy function defined in 
\cite{carlip1} for ${\cal H}^{+}$ only. We here prescribed this entropy function is valid for ${\cal H}^{-}$ as well.}
which is followed from the CFT~\cite{psm} to be
\begin{eqnarray}
{\cal S}_{\pm}(\beta_{\pm})  &=& a\beta_{\pm}+\frac{b}{\beta_{\pm}}   ~. \label{eq4}
\end{eqnarray}
Suppose, we choose the entropy function~\cite{psm} in more general form (because this form admits of a saddle point)  as 
\begin{eqnarray}
{\cal S}_{\pm}(\beta_{\pm})  &=& a\beta_{\pm}^{m}+\frac{b}{\beta_{\pm}^{n}}   ~ \label{eq5}
\end{eqnarray}
where $m,n,a,b>0$. When we have considered the special case which is dictated by CFT then $m=n=1$. 

Now the above function has an extremum value at 
\begin{eqnarray}
\beta_{0, \pm}  &=& \left(\frac{nb}{ma} \right)^{\frac{1}{m+n}}\equiv \frac{1}{T_{\pm}}   ~. \label{eq6}
\end{eqnarray}
Expanding around $\beta_{0, \pm}$ and by evaluating second order derivative one obtains
$$
{\cal S}_{\pm}(\beta_{\pm}) = \alpha \left(a^n b^m \right)^{\frac{1}{m+n}}+
$$
\begin{eqnarray}
\frac{\gamma}{2} \left(a^{n+2} b^{m-2}\right)^{\frac{1}{m+n}} \left(\beta_{\pm}-\beta_{0, \pm}\right)^2 +...  
~ \label{eq7}
\end{eqnarray}
where 
\begin{eqnarray}
\alpha  &=& \left(\frac{n}{m} \right)^{\frac{m}{m+n}}+\left(\frac{m}{n} \right)^{\frac{n}{m+n}},  ~ \label{eq8}
\end{eqnarray}
and 
\begin{eqnarray}
\gamma  &=& \left(m+n\right)m^{\left(\frac{n+2}{m+n}\right)}+n^{\left(\frac{m-2}{m+n}\right)}  ~. \label{eq9}
\end{eqnarray}
We also derived in~\cite{grg1} close to the equilibirium and at the inverse temperature $\beta_{\pm}=\beta_{0, \pm}$, the 
entropy function of ${\cal H}^{\pm}$  as 
\begin{eqnarray}
S_{\pm}(\beta_{\pm}) &=&  S_{0, \pm}+\frac{1}{2} (\beta_{\pm}-\beta_{0, \pm})^2 S_{0, \pm}'' + ...  ~\label{eq10}
\end{eqnarray}
where $S_{0, \pm}: =S_{\pm}(\beta_{0, \pm})$ and $S_{0, \pm}''=\frac{\partial^2 S_{\pm}(\beta_{\pm})}{\partial \beta_{\pm}^2}$
at $\beta_{\pm}=\beta_{0, \pm}$.
Comparing Eq. (\ref{eq7}) and Eq. (\ref{eq10}), we find
\begin{eqnarray}
S_{0, \pm}  &=& \alpha \left(a^n b^m \right)^{\frac{1}{m+n}},    ~ \label{eq11}
\end{eqnarray}
and
\begin{eqnarray}
S_{0, \pm}''  &=&  \gamma \left(a^{n+2} b^{m-2}\right)^{\frac{1}{m+n}}  ~. \label{eq12}
\end{eqnarray}
Inverting these equations one can find $a$ and $b$ in terms of  $S_{0, \pm}$ and $S_{0, \pm}''$ 
\begin{eqnarray}
a &=& \left(\frac{\alpha^{\frac{m-2}{2}}}{\gamma^{\frac{m}{2}}}\right) \left[\frac{(S_{0, \pm}'')^{\frac{m}{2}}}
{(S_{0, \pm})^{\frac{m}{2}}}\right],  ~ \label{eq13}
\end{eqnarray}
and 
\begin{eqnarray}
b &=& \left[\frac{\gamma}{\alpha^{\frac{n+2}{2}}} \right]^{\frac{n}{2}}  \left[\frac{(S_{0, \pm}'')^{\frac{n+2}{2}}}
{(S_{0, \pm})^{\frac{n}{2}}}\right]  ~. \label{eq14}
\end{eqnarray}
Putting these values in Eq. (\ref{eq6}), we get 
\begin{eqnarray}
\beta_{0, \pm}  &=& \frac{1}{T_{\pm}} =\left(\frac{n}{m} \right)^{\frac{1}{m+n}}
\sqrt{\frac{\gamma S_{0, \pm}}{\alpha S_{0, \pm}''}} ~. \label{eq15}
\end{eqnarray}
Now inverting $S_{0, \pm}''$ in terms of $S_{0, \pm}$ and $T_{\pm}$, one obtains 
\begin{eqnarray}
S_{0, \pm}''  &=&  \left[\left(\frac{\gamma}{\alpha}\right)\left(\frac{n}{m}\right)^{\frac{2}{m+n}}\right]
S_{0, \pm} T_{\pm}^2 ~. \label{eq16}
\end{eqnarray}
Substituting these values in Eq. (\ref{eq3}), we have 
\begin{eqnarray}
{\cal S}_{\pm}  &=&  \ln \rho_{\pm} ={\cal S}_{0, \pm}-\frac{1}{2} \ln \left| T_{\pm}^2 S_{0, \pm} \right|+...  ~. \label{eq17}
\end{eqnarray}
This is in fact the generic formula for leading order correction to Bekenstein-Hawking formula. It should be noted that the 
formula is indeed independent of $a$ and $b$. \emph{What is new in this formula is that one could calculate the logarithmic 
correction to the Bekenstein-Hawking entropy of ${\cal H}^{\pm}$ without knowing the values of any specific heat of the BH 
but only knowing the values of $T_{\pm}$ of ${\cal H}^{\pm}$ and $S_{0, \pm}$  for the said BH}.  

Therefore the product becomes
$$
{\cal S}_{+} {\cal S}_{-}  =
{\cal S}_{0, +}{\cal S}_{0, -}-\frac{1}{2}\left[{\cal S}_{0, +} \ln \left| T_{-}^2 {\cal S}_{0,-}\right|+{\cal S}_{0,-}
\ln \left| T_{+}^2 {\cal S}_{0,+}\right|\right]
$$
\begin{eqnarray}
+\frac{1}{4} \ln \left| T_{+}^2 {\cal S}_{0,+} \right| \ln \left|T_{-}^2 {\cal S}_{0,-}\right| +...~. \label{eqq}
\end{eqnarray}
We have already argued the implication of this formula in~\cite{grg1} that when we take the first order correction, it 
indicates that the product is always dependent on mass parameter. Therefore the theorem of Ansorg-Hennig~\cite{ah09} ``The area 
product formula of ${\cal H}^{\pm}$ is independent of mass'' is no longer true when we have taken into consideration the 
leading order logarithmic correction. 

For completeness, we further compute the logarithmic correction of \emph{entropy sum}, \emph{entropy minus} and 
\emph{entropy division} using Eq. (\ref{eq17}). They are
\begin{eqnarray}
{\cal S}_{+}+ {\cal S}_{-} &=& \ln \rho_{+}+\ln \rho_{-}\nonumber\\
&=& \left({\cal S}_{0, +}+{\cal S}_{0, -}\right)-\frac{1}{2}\left[ \ln \left| T_{+}^2 {\cal S}_{0,+}\right|+
\ln \left| T_{-}^2 {\cal S}_{0,-}\right|\right] +...  \nonumber\\ 
{\cal S}_{+}- {\cal S}_{-}  &=& \ln \rho_{+}-\ln \rho_{-}\nonumber\\
&=&\left({\cal S}_{0, +}-{\cal S}_{0, -}\right)-\frac{1}{2}\left[ \ln \left| T_{+}^2 {\cal S}_{0,+}\right|-
\ln \left| T_{-}^2 {\cal S}_{0,-}\right|\right] +...  \nonumber\\
\label{en2}
\end{eqnarray}
and 
\begin{eqnarray}
\frac{ {\cal S}_{+}}{ {\cal S}_{-}} &=& \frac{\ln \rho_{+}}{\ln \rho_{-}}
=\frac{{\cal S}_{0, +}-\frac{1}{2} \ln \left| T_{+}^2 S_{0, +} \right|+...}
{{\cal S}_{0, -}-\frac{1}{2} \ln \left| T_{-}^2 S_{0, -} \right|+...}   ~. \label{en3}
\end{eqnarray}
These equations  indicate that they are all mass-independent hence they are not \emph{universal} in this 
sense.

\section{Examples}

Now we apply this formula for specific BHs in order to calculate the logarithm correction to the 
Bekenstein-Hawking entropy of ${\cal H}^{\pm}$. First we take the four dimensional Reissner 
Nordstr{\o}m (RN) BH.

\emph{Example 1}\\
\emph{RN BH:}\\

The  BH entropy and BH temperature~\cite{davies} becomes
\begin{eqnarray}
{\cal S}_{0, \pm} &=& \pi r_{\pm}^2, \\
T_{\pm} &=& \frac{r_{\pm}-r_{\mp}}{4\pi r_{\pm}^2}  ~\label{eq18}
\end{eqnarray}
where $r_{\pm}=M \pm \sqrt{M^2-Q^2}$, $M$ and $Q$ are mass and charge of BH respectively.

Therefore, the entropy correction is given by
\begin{eqnarray}
{\cal S}_{\pm}  &=&  \ln \rho_{\pm} ={\cal S}_{0, \pm}-\frac{1}{2} \ln \left|\frac{(r_{\pm}-r_{\mp})^2}{16 \pi r_{\pm}^2}\right|
+...  ~. \label{entRn}
\end{eqnarray}
We can conclude that the product includes the mass term so it is not \emph{universal}.

The second example we take is the Kehagias-Sfetsos (KS) BH~\cite{ks09} in Ho\v{r}ava-Lifshitz gravity~
\cite{ph9a,ph9b,ph9c}.

\emph{Example 2}\\
\emph{KS BH:}\\

The  entropy   for KS BH~\cite{horava15} should read 
\begin{eqnarray}
{\cal S}_{0, \pm} &=& \pi r_{\pm}^2,
\end{eqnarray}
and 
\begin{eqnarray}
r_{\pm} &=& M \pm \sqrt{M^2-\frac{1}{2\omega}},
\end{eqnarray}
where $\omega$ is coupling constant, $r_{+}$  and $r_{-}$  are  EH and  CH respectively. 
The Hawking temperature becomes
\begin{eqnarray}
T_{\pm} &=& \frac{\omega (r_{\pm}- r_{\mp})}{4\pi(1+\omega r_{\pm}^2)} ~.\label{M9}
\end{eqnarray}

Therefore, the entropy correction for KS BH  is given by
\begin{eqnarray}
{\cal S}_{\pm}  &=& {\cal S}_{0, \pm}-\frac{1}{2} \ln \left| 
\frac{\omega^2 r_{\pm}^2(r_{\pm}-r_{\mp})^2}{16 \pi (1+\omega r_{\pm}^2) }\right|+...  ~. \label{c6}
\end{eqnarray}
It implies that  when the logarithmic correction is taken into consideration the entropy product is not 
mass-independent (universal).

Now we take the AdS space. First it should be Schwarzschild-AdS space-time.
\footnote{In the limit $\ell \rightarrow \infty$, one gets the Schwarzschild BH. Here the horizon is at 
$r_{+}=2M$ and ${\cal S}_{0,+}=4\pi M^2$.  Thus one obtains the log correction as 
${\cal S}_{+}=4\pi M^2+\frac{1}{2} \ln|16\pi|$. Therefore it indicates that the logarithmic correction must be 
mass dependent. Therefore it is not  universal.}

\emph{Example 3}\\
\emph{Schwarzschild-AdS BH:}\\

The only  physical horizon~\cite{mv13} is at
\begin{eqnarray}
{r}_{ +} &=&\frac{2\ell}{\sqrt{3}} sinh \left[\frac{1}{3}sinh^{-1}\left(3\sqrt{3}\frac{M}{\ell}\right)\right] ~.\label{ehd}
\end{eqnarray}
Thus the entropy of ${\cal H}^{+}$ should be 
\begin{eqnarray}
 {\cal S}_{0, +} &=& \pi r_{+}^2, ~\label{pd3}
\end{eqnarray}
where $\Lambda=-\frac{3}{\ell^2}$ is cosmological constant.
The BH temperature reads 
\begin{eqnarray}
T_{+} &=& \frac{1}{4\pi r_{+}} \left(1+ 3\frac{r_{+}^2}{\ell^2} \right) ~.\label{pd4}
\end{eqnarray}
Now the entropy correction formula should read 
\begin{eqnarray}
{\cal S}_{+}  &=& {\cal S}_{0, +}-\frac{1}{2} \ln \left|\frac{\left(1+3\frac{r_{+}^2}{\ell^2} \right)^2}
{16 \pi} \right|+...  
~. \label{sads}
\end{eqnarray}
In fact, in both cases, with logarithmic correction and without logarithmic correction, the entropy depends on the mass 
parameter thus it is not universal and therefore it is not quantized.

Now we  take the RN-AdS case~\cite{mv13}. 

\emph{Example 4}\\
\emph{RN-AdS BH:}\\

The quartic Killing horizon equation becomes
\begin{eqnarray}
 r^4 +\ell^2r^2-2M\ell^2r+Q^2\ell^2 &=& 0 ~.\label{rn1}
\end{eqnarray}
There are at least two real zeros which correspond to two physical horizons namely EH, $r_{+}$ and CH, $r_{-}$. 

The entropy should read 
\begin{eqnarray}
{\cal S}_{0, \pm} &=& \pi r_{\pm}^2 ~.\label{eq19}
\end{eqnarray}
The BH temperature of ${\cal H}^{\pm}$ is given by 
\begin{eqnarray}
T_{\pm} &=& \frac{1}{4\pi r_{\pm}} \left( 3\frac{r_{\pm}^2}{\ell^2}-\frac{Q^2}{r_{\pm}^2}+1\right) ~.\label{rn2}
\end{eqnarray}

Therefore the logarithmic correction becomes
\begin{eqnarray}
{\cal S}_{\pm}  &=& {\cal S}_{0, \pm}-\frac{1}{2} \ln \left|\frac{\left(3\frac{r_{+}^2}{\ell^2}-\frac{Q^2}{r_{+}^2}+1 \right)^2}
{16 \pi} \right|+...  
~. \label{rn4}
\end{eqnarray}
It implies that the product of logarithmic correction is not mass independent.

Now we take the spinning BH.

\emph{Example 5}\\
\emph{Kerr BH:}\\

The  BH entropy and BH temperature~\cite{davies} are 
\begin{eqnarray}
{\cal S}_{0, \pm} &=& \pi \left(r_{\pm}^2+a^2\right), \\
T_{\pm} &=& \frac{r_{\pm}-r_{\mp}}{4\pi \left(r_{\pm}^2+a^2\right)} 
~\label{krt}
\end{eqnarray}
where $r_{\pm}=M \pm \sqrt{M^2-a^2}$, $M$ and $a=\frac{J}{M}$ are mass and spin parameter of the BH respectively.

Now the logarithmic correction is computed to be
\begin{eqnarray}
{\cal S}_{\pm}  &=& {\cal S}_{0, \pm}-\frac{1}{2} \ln \left|\frac{\left(r_{\pm}-r_{\mp}\right)^2}
{16 \pi \left(r_{\pm}^2+a^2\right)}\right|+...  ~. \label{kr4}
\end{eqnarray}
It also indicates that whenever we take the logarithmic correction, the entropy product of ${\cal H}^{\pm}$ 
is not universal. 

Next we take charged rotating BH.

\emph{Example 6}\\
\emph{Kerr-Newman BH:}\\

The  BH entropy and BH temperature~\cite{davies} should read
\begin{eqnarray}
{\cal S}_{0, \pm} &=& \pi \left(r_{\pm}^2+a^2\right), \\
T_{\pm} &=& \frac{r_{\pm}-r_{\mp}}{4\pi \left(r_{\pm}^2+a^2\right)} 
~\label{kn1}
\end{eqnarray}
where $r_{\pm}=M \pm \sqrt{M^2-a^2-Q^2}$, $M$, $a$ and $Q$ correspods to the mass, the spin parameter and the charge of
BH respectively.

The logarithmic correction is derived to be
\begin{eqnarray}
{\cal S}_{\pm}  &=& {\cal S}_{0, \pm}-\frac{1}{2} \ln \left|\frac{\left(r_{\pm}-r_{\mp}\right)^2}
{16 \pi \left(r_{\pm}^2+a^2\right)}\right|+...  ~. \label{kn4}
\end{eqnarray}
Again we observe that when we take logarithmic correction the entropy product of ${\cal H}^{\pm}$ for 
Kerr-Newman BH~\cite{ah09,pp14} is not mass independent. 

\emph{Example 7}\\
\emph{Kerr-Newman AdS BH:}\\

The horizon equation \cite{marco} is given by
\begin{eqnarray}
\Delta_{r}= \left(r^2+a^2 \right)\left(1+\frac{r^2}{\ell^2}\right)-2Mr+Q_{e}^2+Q_{m}^2 =0 ~\label{kk1}
\end{eqnarray}
which implies that the quartic order horizon equation
\begin{eqnarray}
 r^4 +\left(\ell^2+a^2\right)r^2-2M\ell^2r+\left(a^2+Q_{e}^2+Q_{m}^2\right)\ell^2 = 0 ~.\label{kk2}
\end{eqnarray}
This equation has two real zeros which correspond to two physical horizons namely $r_{\pm}$, where $Q_{e}$ and 
$Q_{m}$ are electric and magnetic charge parameters respectively. 
The  BH entropy and BH temperature are
\begin{eqnarray}
{\cal S}_{0, \pm} &=& \frac{\pi \left(r_{\pm}^2+a^2\right)}{\left(1-\frac{a^2}{\ell^2}\right)} \\
T_{\pm} &=& \frac{r_{\pm}\left(1+\frac{a^2}{\ell^2}+3\frac{r_{\pm}^2}{\ell^2}-
\frac{\left(a^2+Q_{e}^2+Q_{m}^2\right)}{r_{\pm}^2} \right)}{4\pi \left(r_{\pm}^2+a^2\right)} 
~.\label{kk3}
\end{eqnarray}

The logarithmic correction for KN-AdS BH should read
\begin{eqnarray}
{\cal S}_{\pm} = {\cal S}_{0, \pm}-\frac{1}{2} \ln \left|\frac{r_{\pm}^2 \left(1+\frac{a^2}{\ell^2}+3\frac{r_{\pm}^2}{\ell^2}-
\frac{\left(a^2+Q_{e}^2+Q_{m}^2\right)}{r_{\pm}^2} \right)^2}{16 \pi\left(1-\frac{a^2}{\ell^2}\right) 
\left(r_{\pm}^2+a^2\right)}\right|+... \nonumber\\
~. \label{kk4}
\end{eqnarray}

\emph{Example 8}\\
\emph{Non-rotating BTZ BH:}\\

The BH horizon is at $r_{+}=\sqrt{8G_{3}M \ell}$. $G_{3}$ is 3D Newtonian constant. The BH entropy of 
${\cal H}^{+}$ for BTZ BH is  
\begin{eqnarray}
 {\cal S}_{0, +} &=& \frac{2\pi r_{+}}{4G_{3}} ~\label{bt1}
\end{eqnarray}
The BH temperature is  
\begin{eqnarray}
T_{+} &=& \frac{r_{+}}{2\pi \ell^2} ~\label{bt2}
\end{eqnarray}
where $\Lambda=-\frac{3}{\ell^2}$ is cosmological constant.
Thus the BH entropy correction for BTZ BH is 
\begin{eqnarray}
{\cal S}_{+}  &=& {\cal S}_{0, +}-\frac{1}{2} \ln \left|\frac{r_{+}^3}{8 \pi G_{3} \ell^4 } \right|+...  
~. \label{bt3}
\end{eqnarray}
In fact, it is isolated case and there is only one horizon therefore here both the log correction of entropy and without 
log correction term is mass dependent.

\emph{Example 9}\\
\emph{Rotating BTZ BH:}\\

The BH horizons for rotating BTZ BH  \cite{btz92,jetpl} are given by
\begin{equation} 
r_{\pm}= \sqrt{4G_{3}{\cal M} \ell^2\left(1\pm \sqrt{1-\frac{J^2}{{\cal M}^2 \ell^2}} \right)}
 .~\label{btr1}
\end{equation}
The BH entropy of ${\cal H}^{\pm}$  is  
\begin{eqnarray}
 {\cal S}_{0, \pm} &=& \frac{2\pi r_{\pm}}{4G_{3}}, ~\label{btr2}
\end{eqnarray}
and the Hawking temperature of ${\cal H}^{\pm}$ should read
\begin{eqnarray}
T_{\pm} &=& \frac{r_{\pm}^2-r_{\mp}^2}{2\pi \ell^2 r_{\pm}} .~\label{btr3}
\end{eqnarray}

Therefore the BH entropy correction is calculated to be  
\begin{eqnarray}
{\cal S}_{\pm}  &=& {\cal S}_{0, \pm}-\frac{1}{2} \ln \left|\frac{(r_{\pm}^2-r_{\mp}^2)^2}{8 \pi G_{3} \ell^4 r_{\pm} } \right|+...  
~. \label{btr4}
\end{eqnarray}
It is clear from the calculation that the product depends on the mass parameter.

\emph{Example 10}\\
\emph{Charged Dilaton BH:}\\

The horizons of charged dilation BH \cite{ahep} is at
\begin{eqnarray}
r_{+} &=& M+ \sqrt{M^{2}-\left(\frac{2n}{1+n}\right)Q^{2}} ~.\label{cde} \\
r_{-} &=& \frac{1}{n}\left[M+ \sqrt{M^{2}-\left(\frac{2n}{1+n}\right)Q^{2}}\right]
~.\label{cdc}
\end{eqnarray}
where $n$ is given by
\begin{eqnarray}
n &=& \frac{1-a^{2}}{1+a^{2}}.
\end{eqnarray}

The  BH temperature of ${\cal H}^\pm$ reads
\begin{eqnarray}
T_{+} &=& \frac{1}{4\pi r_{+}}\left(\frac{r_{+}-r_{-}}{r_{+}}\right)^{n} \\
\mbox{and} \nonumber\\
T_{-}&=& 0
\end{eqnarray}

The entropy for both the horizons are
\begin{eqnarray}
{\cal S}_{0, +} &=&  \pi r_{+}^2 \left(\frac{r_{+}-r_{-}}{r_{+}}\right)^{1-n} \,\, \mbox{and}\,\,  
{\cal S}_{0, -}=  0\,\,~.\label{pcd}
\end{eqnarray}

Thus the BH entropy correction should be for  ${\cal H}^{+}$
\begin{eqnarray}
{\cal S}_{+}  &=& {\cal S}_{0, +}-\frac{1}{2} \ln \left|\frac{\left(\frac{r_{+}-r_{-}}{r_{+}}\right)^{1+n}}{16 \pi } \right|+...  
~. \label{cd1}
\end{eqnarray}
and the entropy correction for  ${\cal H}^{-}$ is 
\begin{eqnarray}
{\cal S}_{+}  &=& 0  
~. \label{cd2}
\end{eqnarray}
This is an interesting case because the entropy product of ${\cal H}^\pm$ and the entropy product with logarithmic correction 
both go to zero value. The logarithmic correction survives for ${\cal H}^{+}$ only when we have taken  into account the 
logarithmic correction for ${\cal H}^{-}$ it breaks down and should therefore the product also breaks down.

\emph{Example 11}\\
\emph{Kerr-Sen  BH:}\\

The horizons for Kerr-Sen \cite{as92,ppsen} BH is  situated at  
\begin{eqnarray}
r_{\pm} &=& \left(M-\frac{Q^2}{2M}\right)\,\, \pm \sqrt{\left(M-\frac{Q^2}{2M}\right)^{2}-a^2}
~.\label{s1}
\end{eqnarray}

The  BH entropy and BH temperature for Sen BH are
\begin{eqnarray}
{\cal S}_{0, \pm} &=& 2 \pi M r_{\pm} \\
T_{\pm} &=& \frac{r_{\pm}-r_{\mp}}{8\pi M r_{\pm}} 
~.\label{s2}
\end{eqnarray}

Therefore the logarithmic correction is calculated to be 
\begin{eqnarray}
{\cal S}_{\pm}  &=& {\cal S}_{0, \pm}-\frac{1}{2} \ln \left|\frac{\left(r_{\pm}-r_{\mp}\right)^2}
{32 \pi M r_{\pm}}\right|+...  ~. \label{s3}
\end{eqnarray}

\emph{Example 12}\\
\emph{Sultana-Dyer  BH:}\\
This is an example of a dynamical cosmological BH \cite{sultana}. The horizon is located at $r_{+}=2m$. Where $m$ is the mass of
BH. 

The entropy \footnote{The surface area at $t=0$ indicates that the cosmological BH is formed initially from Big-Bang singularity. 
$t^4$ is the conformal factor.} 
and temperature of this  BH are
\begin{eqnarray}
{\cal S}_{0, +} &=& \frac{{\cal A}_{+}}{4}= \pi t^4 r_{+} \\
T_{+} &=& \frac{1}{4\pi t^2 r_{+}} 
~.\label{sd}
\end{eqnarray}

The logarithmic correction is found to be 
\begin{eqnarray}
{\cal S}_{+}  &=& {\cal S}_{0, +}-\frac{1}{2} \ln \left|\frac{1}{16 \pi} \right|+...  ~. \label{sd3}
\end{eqnarray}
The interesting fact in this case that we have found the entropy correction term  in the logarithmic correction term 
is mass-independent whereas without logarithmic term is mass dependent. 

\emph{Example 13}\\
\emph{Charged BHs in $f(R)$  gravity:}\\

The $f(R)$ gravity \cite{moon,castro} is interesting because it is equivalent to Einstein gravity coupled to matter, where $f(R)$ 
is an arbitrary function of the scalar curvature. The horizon function at the constant scalar curvature $R=R_{0}$ is given by 
\begin{eqnarray}
{\cal N}(r) &=&  1-\frac{2\mu}{r}+\frac{q^2}{\alpha r^2}-\frac{R_{0}}{12}r^2=0 ~ \label{fr}
\end{eqnarray}
where $\alpha=1+f'(R_{0})$. The quantities $\mu$ and $q$ are related to the $M$ (ADM mass) and $Q$ (electric charge) in 
this gravity becomes
\begin{eqnarray}
M = \mu \alpha ,\,\,\,\, Q=\frac{q}{\sqrt{2\alpha}}~. \label{fr1}
\end{eqnarray}
The entropy for all the horizons is
\begin{eqnarray}
{\cal S}_{0, i}  &=& \pi \alpha  r_{i}^2, ~\label{fr2}
\end{eqnarray}
and the BH temperature  should read 
\begin{eqnarray}
T_{i} &=&  \frac{1}{4\pi r_{i}} \left(1-\frac{q^2}{\alpha r_{i}^2} -\frac{R_{0}}{4}r_{i}^2 \right)
~.\label{fr3}
\end{eqnarray} 
The logarithmic correction of entropy  becomes
\begin{eqnarray}
{\cal S}_{i}  &=& {\cal S}_{0, i}-\frac{1}{2} \ln \left|\frac{\alpha \left(1-\frac{q^2}{\alpha r_{i}^2} 
-\frac{R_{0}}{4}r_{i}^2 \right)^2}{16 \pi r_{i}^2} \right|+...  ~. \label{fr4}
\end{eqnarray}

\emph{Example 14}\\
\emph{5D Gauss Bonnet BH:}\\
The horizon radii for 5D Gauss-Bonnet BH~\cite{castro} are located at
\begin{eqnarray}
r_{\pm} &=& \frac{1}{\sqrt{2}} \sqrt{\left(2\mu-\alpha_{d}\right)\,\, \pm \sqrt{\left(2\mu-\alpha_{d}\right)^{2}-4q^2}}
~.\label{gb1}
\end{eqnarray}
where $\mu=\frac{4M}{3\pi}$ and $q=(\frac{4}{\pi})^{\frac{2}{3}}Q$. 

The entropy of  ${\cal H}^\pm$ is
\begin{eqnarray}
{\cal S}_{0, \pm}  &=& \frac{\pi^2r_{\pm}^3}{2} \left(1+\frac{6\alpha_{d}}{r_{\pm}^2} \right) ~.\label{gb2}
\end{eqnarray}

The BH temperature of  ${\cal H}^\pm$ reads
\begin{eqnarray}
T_{\pm} &=&  \frac{r_{\pm}^2-r_{\mp}^2}{2\pi r_{\pm} \left(r_{\pm}^2+2\alpha_{d}\right)}
~,\label{gb3}
\end{eqnarray} 
where $\alpha_{d}=(d-3)(d-4)\alpha=2\alpha$ and $\alpha_{d}$ is Gauss-Bonnet coupling constant.

The logarithmic correction of entropy ${\cal H}^\pm$ becomes
\begin{eqnarray}
{\cal S}_{\pm}  &=& {\cal S}_{0, \pm}-\frac{1}{2} \ln \left|\frac{r_{\pm} \left(1+\frac{6\alpha_{d}}{r_{\pm}^2}\right)
\left(r_{\pm}^2-r_{\mp}^2 \right)^2}{8 \left(r_{\pm}^2+2\alpha_{d}\right)^2} \right|+...  ~. \label{gb5}
\end{eqnarray}
It follows from the several examples that when the logarithmic correction is considered the entropy product 
formula is \emph{not mass-independent (universal)} and therefore it \emph{is not quantized}. 

To sum up, we computed the general logarithmic corrections to the BH entropy product formula of inner horizon and outer 
horizon by taking into consideration effects of statistical  quantum fluctuations around the thermal equilibirium and also 
via CFT. We showed followed by our earlier work~\cite{grg1} that whenever we take the first order logarithmic correction 
to the entropy product formula, it is not universal and  also it can not be quantized. What is new in this 
work is that when we have choosen  the exact entropy function followed by CFT and by taking the effects of 
quantum fluctuations, the logarithmic correction formula of ${\cal H}^\pm$ should depends solely on the value 
of BH temperature of ${\cal H}^\pm$  and BH entropy  of ${\cal H}^\pm$ at the  thermal equilibirium.



\end{document}